\documentclass{elsart}
\usepackage{amssymb}
\usepackage{graphicx,color}% Include figure files

\begin {document}

\begin{frontmatter}
%\preprint{LBNL-63707}
{\hfill LBNL-}

\title{Multiple parton scattering in nuclei: gauge invariance}

\author[shanda]{Zuo-tang Liang},
\author[lbnl,shanda]{Xin-Nian Wang} and
\author[shanda,lbnl]{Jian Zhou}

\address[shanda]{Department of Physics, Shandong University, Jinan,
Shandong 250100, China}
\address[lbnl]{Nuclear Science Division, MS 70R0319,
Lawrence Berkeley National Laboratory, Berkeley, California 94720}

\begin{abstract}
Within the framework of a generalized collinear factorization for multiple parton
scattering in nuclear medium, twist-4 contributions to DIS off a large
nucleus can be factorized as a convolution of hard parts and two-parton
correlation functions. The hard parts for the quark scattering in the 
light-cone gauge correspond to interaction with a transverse (physical) 
gluon in the target, while they are given by the second derivative of the cross
section with a longitudinal gluon in the covariant gauge. We provide a general 
proof of the equivalence of the hard parts in the light-cone and covariant gauge.
We further demonstrate the equivalence in calculations of twist-4
contributions to semi-inclusive cross section of DIS in both lowest order
and next leading order. Calculations of the nuclear transverse momentum broadening 
of the struck quark in the light-cone and covariant gauge are also discussed.
\end{abstract}

\begin{keyword}
\PACS{...}
\end{keyword}

\end{frontmatter}

\section {Introduction}

In the study of nuclear matter and quark-gluon plasma in high
energy lepton-nucleus, hadron-nucleus and nucleus-nucleus collisions, hard
processes such as jet and high transverse momentum hadrons production are
useful tools as their initial production rates can be calculated within
perturbative QCD (pQCD). Modification of the final jet or hadron spectra
known as jet quenching \cite{wg90} due to further interaction between
the energetic partons and the
nuclear or hot medium can be used to probe the properties of medium such as
parton correlation or gluon density \cite{ww00}. Indeed, strong jet
quenching has been observed both in high-energy lepton-nucleus \cite{hermes}
and nucleus-nucleus collisions \cite{phenix,star}.

Current phenomenological studies of experimental data on jet quenching to
extract properties of the nuclear matter or hot medium rely
on pQCD calculations of the parton energy loss or modification of
the parton fragmentation functions due to gluon radiation induced by
multiple scattering during the parton
propagation. Among many theoretical studies \cite{Gyulassy:2003mc,Kovner:2003zj},
twist expansion approach \cite{gw-twist,owz} was applied to medium modification
of the fragmentation functions and parton energy loss in both nuclei in
deeply inelastic scattering (DIS) and hot QCD matter in heavy-ion
collisions. Such an approach is based on the generalized factorization
framework for multiple parton scattering in nuclear medium first
developed by Luo, Qiu and Sterman (LQS) \cite{LQS}. Within this framework,
LQS proved that the leading twist-4 contribution from
multiple parton scattering in a large nucleus can be factorized
as a convolution of hard parts and two-parton correlation
functions. Such parton correlation functions have leading contributions
that are proportional to the nuclear size $R_A \sim A^{1/3}$ and therefore 
the twist-4 contributions are enhanced by the nuclear medium as compared 
to the same contribution in nucleon collisions.

The twist expansion technique has been used to calculate
the effect of multiple parton scattering in dijet and photon
production in DIS off nuclear targets \cite{LQS2,Guo:1995zk} and
Drell-Yan (DY) dilepton spectra in $p+A$ collisions \cite{Guo:1997it,Fries:1999jj}.
In these calculations, processes involving secondary
scattering with a soft gluon (soft scattering) are treated
in the framework of collinear expansion in a covariant gauge
while processes of secondary scattering with a hard gluon
(double hard scattering) are calculated directly in the light-cone
gauge with collinear approximation (neglecting the transverse
momentum of the initial gluons from the nucleus). In such
separate treatments of soft and double hard scattering, it is
difficult to include the interference between the two type of
processes.
For large values of the final transverse momentum of the dijets,
photons or DY dilepton, one nevertheless can neglect the
interference effects since the formation time of the particle
production, which dictates the interference, is much smaller than the
nuclear size. In the investigation of nuclear modification of parton
fragmentation functions and induced parton energy loss \cite{gw-twist,owz},
it is critical to consider soft gluon bremsstrahlung. Since the
formation time of soft gluons can be larger or comparable to
the nuclear size, one has to include the interference between
soft and double hard scattering which is equivalent to the
Landau-Pomeranchuck-Migdal (LPM) interference \cite{lpm}.
In this case, both soft and double hard processes and their interference
terms should be calculated within the general framework of
collinear expansion in the same gauge.

In a covariant gauge, the longitudinal component of the gauge
field $A^+$ is considered large. One usually makes a collinear
expansion of the hard part of quark and (longitudinal) gluon
interaction. These hard parts do not correspond to any physical
quark-gluon scattering. If one expands the hard parts in the
transverse momentum of the longitudinal gluon field, the collinear
terms are directly related to the hard parts of the vacuum diagram
without final gluon interaction. They only contribute to the gauge
link of the initial quark distribution function of the vacuum diagram.
The higher order terms in the collinear expansion will contribute to 
higher twist cross section involving two-parton correlation functions
and derivatives of the corresponding hard parts.
In a light-cone gauge $A^+=0$, contributions to the higher
twist cross section can be written directly as a convolution
of the two-parton correlation function and the hard part
of collinear quark and (transverse or physical) gluon interaction.
Since the two-parton correlation functions can be expressed in a
gauge-invariant form, the collinear hard part in the light-cone
gauge should be equivalent to the derivatives of the hard part
in the covariant gauge and the final higher-twist cross section
should be gauge independent.

In this paper, we first provide a general proof of the equivalence
of calculations in both light-cone and covariant gauge of the leading
twist-4 contributions to  multiple parton scattering processes at
any order of $\alpha_s$ in DIS off a large nucleus. We then
demonstrate the equivalence with explicit calculations
of lowest order semi-inclusive DIS cross section, nuclear transverse
momentum broadening and induced gluon radiation
via secondary parton scattering in DIS in both light-cone and
covariant gauge.

%%%%%%%%%%%%%%%%%%%%%%%%%%%%%%%%%%%%%%%%%%%%%%%%%%%%%%%%%%%%%%%%%%%%%
%%%%%%%%%%%%%%%%%%%%%%%%%%%%%%%%%%%%%%%%%%%%%%%%%%%

\section {Generalized collinear factorization}

For simplification, we only consider the
semi-inclusive differential cross section for the final quark and
do not consider quark fragmentation. The differential cross section for
$e(l)+A(p)\rightarrow e(l^\prime)+q(\ell)+X$ can be written as
\begin{equation}
d\sigma=\frac{\alpha_{em}^2e_q^2}{\pi sQ^4}L^{\mu\nu}(l,l')
\frac{dW_{\mu\nu}(q,p,\ell)}{d^2\ell_\perp}
\frac{d^3l'}{E_{l'}}d^2\ell_\perp,
\label{sidis}
\end{equation}
where $L^{\mu\nu}(l,l')=4[l^\mu{l'}^\nu+l^\nu{l'}^\mu-(l\cdot l')g^{\mu\nu}]$
is the leptonic tensor, $p=[p^+,0,\vec{0}_\perp]$ is the longitudinal momentum
per target nucleon, $q=[-Q^2/2q^-,q^-,0]$ the 4-momentum of the virtual photon,
$s=(l+p)^2$ and $\ell$ is the 4-momentum of the outgoing quark.

In a large nucleus or hot QCD matter in heavy-ion collisions,
a produced initial parton may experience additional scattering
with other partons from the medium. The additional scattering
may further induce additional gluon radiation and cause the parton to
lose energy. Such multiple scattering and induced gluon radiation will
effectively lead to modification of the parton fragmentation functions
in a medium, photon and dilepton production. These additional
contributions from multiple scattering are higher-twist
contributions and are always power-suppressed. They generally involve
high-twist matrix elements of
parton correlation of the medium. For multiple scattering in cold
nuclear matter, two-parton correlations can involve partons from
different nucleons in the nucleus, they are proportional to the
size of the nucleus and thus are enhanced by a nuclear factor $A^{1/3}$
as compared to two-parton correlations in a nucleon.

\begin{figure}
\begin{center}
\includegraphics[width=3.0in,height=2.0in]{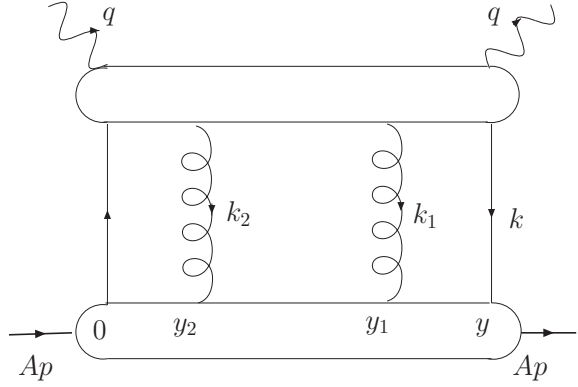}
\end{center}
\caption{Two-gluon interaction in DIS}
\label{fig1}
\end{figure}

In general, the leading (or nuclear enhanced) twist-four
contributions at all orders in $\alpha_{\rm s}$ to the hadronic
tensor of DIS off a nucleus involving a secondary scattering with
another gluon (see Fig.~\ref{fig1}) can be written as
\begin{eqnarray}
W^{(2)}_{\mu \nu} &= &\int \frac{d^4 k}{(2 \pi)^4} \int
\frac{d^4 k_1}{(2 \pi)^4} \int \frac{d^4 k_2}{(2
\pi)^4}    \int d^4 y  \int d^4 y_1  \int d^4 y_2 \,
e^{iky+ik_1y_1+ik_2 y_2}
 \nonumber\\
&\times & \frac{1}{p\cdot n}
{\rm Tr} [\hat{H}^{\rho \sigma}_{ \mu \nu} (k,k_1,k_2)\frac{p\!\!/}{2} ]
\langle A \mid
\bar{\psi}(0) \frac{n\!\!\!/}{2}
A_{\rho}(y_1) A_{\sigma}(y_2)\psi(y) \mid A\rangle
\end{eqnarray}
where $\hat{H}^{\rho \sigma }_{\mu \nu} (k,k_1,k_2)$ is the perturbative
hard part of the multiple parton scattering at any order of the strong
coupling constant, $n=[0,1,\vec{0}_\perp]$ and $\bar n=[1,0,\vec 0_\perp]$
are unit four-vectors.
Here we have suppressed the color index. Summations over color
indices of the field operators in the matrix element and average over the
color indices of the initial state partons in the hard part
are understood. To simplify the notation,
we will sometimes suppress the polarization indices associated with the
photon interaction when they are not necessary,
$\hat{H}^{\rho \sigma} (k,k_1,k_2)\equiv
\hat{H}^{\rho \sigma}_{\mu \nu} (k,k_1,k_2)$. In the above
expression, one can always assume collinear approximation
for the quark fields, $k=xp$. The transverse momentum of the quark
will only contribute to high twist contributions that are not
nuclear enhanced. We can further neglect the minus-components
of the gluon's momentum, which also contribute to higher-twist
terms that are not nuclear enhanced. One can therefore express
the gluon's 4-momentum as
$k_1^\rho=x_1p^\rho +k^\rho_{1\bot}$,
$k_2^\rho=x_2p^\rho +k^\rho_{2\bot}$.
One can also decompose gluon fields as $A^\rho= A^\rho _\bot + A ^\rho _+ +A^\rho _-$.
Here $ A^\rho_\bot=A_\sigma d^{\rho \sigma}$,
$A^\rho_+ =p^\rho A \cdot n/p \cdot n$,
$A^\rho _-= n^\rho A \cdot p/p \cdot n $,
and $d_{\rho \sigma}= -g_{\rho \sigma}+\bar{n}_\rho n_\sigma
+n_\rho \bar{n}_\sigma$ is the transverse projection
($d_{\rho \sigma}A^\rho B^\sigma=\vec A_\perp \cdot \vec B_\perp$).

\subsection{Light-cone gauge}

Let us first consider the above hadronic tensor in the light-cone
gauge , $A^\rho_+=0$. In this case, $A^\rho_-$ component contributes
only to higher twist (beyond twist-4). Therefore, the matrix elements
of the non-perturbative part of the hadronic tensor only involve
one term;
\begin{eqnarray}
&&\langle A \mid\bar{\psi}(0) n\!\!\!/ A_\bot ^{\rho}(y_1)
A_\bot ^{\sigma}(y_2)\psi(y) \mid A \rangle \nonumber \\
&&\hspace{1.0in}= -\frac{1}{2} d^{\rho \sigma }
\langle A \mid\bar{\psi}(0) n\!\!\!/ \vec A_\bot (y_1)\cdot A_\bot (y_2)
\psi(y) \mid A \rangle ,
\end{eqnarray}
for unpolarized nuclei.

Taking collinear approximation (neglecting gluons' transverse momentum),
one can carry out all the integrations except $y_{i}^-$ and $k_i^+=x_ip^+$.
One can further integrate by parts in $y_1^-, y_2^-$ to convert
$-x_1p^+\vec A_\bot (y_1^-) \cdot x_2p^+\vec A_\bot (y_2^-)$ into
gluon field strength,
 $\partial_+ \vec A_\bot (y_1) \cdot \vec \partial_+ A_\bot (y_2)
=\vec F_{+\bot}(y_1^-)\cdot \vec F_{+\bot}(y_2^-)$ (Note that
$\vec F_{+\bot}=\partial _+\vec A_\bot $ in the light-cone gauge, $A^+ = 0$).
The hadronic tensor in the light-cone gauge can be reduced to
\begin{eqnarray}
 W^{(2)}&=& \int
 \frac{dy^-}{2\pi}\frac{dy^-_1}{2\pi}\frac{dy^-_2}{2\pi}\int dx dx_1dx_2
e^{ix_1p^+y_1^- + ix_2p^+y_2^- + ixp^+y^-}
\nonumber\\
& \times &\langle A \mid \bar{\psi}(0)\frac{n\!\!\!/}{2}
\vec F_{+\bot}(y^-_1)\cdot \vec F_{+\bot} (y^-_2)
\psi(y^-)\mid A\rangle \nonumber\\
&\times& {\rm Tr} [
\hat{H}^{\rho \sigma}(xp,x_1p,x_2p)
\frac{p\!\!/d_{\rho\sigma}}{4x_1 x_2}],
\label{eq:lgw}
\end{eqnarray}
where we have suppressed both the $+$ and transverse components
of the coordinates when they are set to zero, $y^+=0$, $\vec y_\perp=0$,
in variables of the field operators.

The above formula is normally used to calculate contributions from
double hard scattering where gluons carry finite momentum fraction
$x_1$ and $x_2$. However, in soft scattering processes, the
momentum fractions $x_1$ and $x_2$ go to zero. One should then
choose a regularization prescription on these light-cone poles
corresponding to specific boundary conditions for the gluon
field \cite{Boer:1997bw}. In principal, all the prescriptions are
equivalent for the complete set of diagrams. Alternatively,
one can calculate both types of multiple parton scattering
and their interferences in the covariant gauge as was outlined
by LQS \cite{LQS}.

\subsection{Covariant gauge}

In the covariant gauge, both the transverse and plus components of the
gluon fields contribute to the same twist operators while the minus component
contributes only to higher twist terms. The matrix elements of the
non-perturbative part can therefore be expanded into four terms,
\begin{eqnarray}
\langle A \mid
 \bar{\psi}(0)[A_+^\rho(y_1^-)A_+^\sigma(y^-_2)
&+&A_+^\rho(y_1^-)A_\bot^\sigma(y_2^-)
 +A_\bot^\rho(y_1^-)A_+^\sigma(y_2^-) \nonumber \\
&+&A_\bot^\rho(y_1^-)A_\bot^\sigma(y_2^-)]\psi(y^-)
 \mid A\rangle . \nonumber
\end{eqnarray}
To isolate the leading terms of the twist-4 contribution with nuclear
enhancement, one can follow LQS \cite{LQS} to make the following
collinear expansion of the perturbative hard part in terms of the
gluon transverse momentum~\footnote{Note that when only the relative
transverse momentum $\vec k_\bot=\vec k_{1\bot}=-\vec k_{2\bot}$ is considered as
in Refs.~\cite{gw-twist},
$\hat{H}(\vec k_{1\bot}-\vec k_{2\bot})=\hat{H}(2\vec k_\bot)$} ,
\begin{eqnarray}
\hat{H}^{\rho \sigma}(k,k_1,k_2)&=&\hat{H}^{\rho \sigma}(k,x_1p,x_2p)
 +\frac{\partial \hat{H}^{\rho \sigma}(k,k_1,x_2p)} {\partial k_1^\alpha}
 \mid _{k_1=x_1p} k_{1\bot}^\alpha
\nonumber\\
&+&\frac{\partial \hat{H}^{\rho \sigma}(k,x_1p,k_2)} {\partial
k_2^\alpha} \mid _{k_2=x_2p} k_{2\bot}^\alpha  \nonumber\\
&+&   \frac{\partial ^2
\hat{H}^{\rho \sigma}(k,k_1,k_2)} {\partial k_1^\alpha
\partial k_2^\beta} \mid _{k_1=x_1p, k_2=x_2p} k_{1\bot}^\alpha k_{2\bot}^\beta
+...
\nonumber\\
\end{eqnarray}
where terms $\partial ^2
\hat{H}^{\rho \sigma}/\partial k_1^{\alpha}\partial k_1^{\beta}$ and
$\partial ^2 \hat{H}^{\rho \sigma}/\partial k_2^{\alpha}\partial k_2^{\beta}$
are not listed as they don't contribute to the nuclear
enhanced twist-4 terms. The first (collinear) term can be reduced to the
eikonal term of the twist-2 contribution making the twist-2 quark
distribution function gauge invariant. The rest of the expansion
will contribute to higher twist terms. Again, after integration by parts
and considering
$\partial_+A_\bot^\rho\partial_+A_\bot^\sigma
=\partial_+\vec A_\bot\cdot\partial_+\vec A_\bot d^{\rho\sigma}/2$,
$\partial_+A_\bot^\rho\partial_\bot^\sigma A_+
=\partial_+\vec A_\bot\cdot \vec\partial_\bot A_+ d^{\rho\sigma}/2$,
$\partial_\bot^\rho A_+\partial_\bot^\sigma A_+
=\vec \partial_\bot A_+\cdot\vec \partial_\bot A_+ d^{\rho\sigma}/2$
in a unpolarized nuclear target, one gets

\begin{eqnarray}
W^{(2)}&=& \frac{1}{8} \int
 \frac{dy^-}{2\pi}\frac{dy^-_1}{2\pi}\frac{dy^-_2}{2\pi}\int dx_1dx_2dx
e^{ix_1p^+y_1^- + ix_2p^+y_2^- + ixp^+y^-}
\nonumber\\
& & \left \{ d^{\alpha\beta}
{\rm Tr}\left[\frac{\partial ^2 \hat{H}_{\rho \sigma}(k,k_1,k_2)}
{\partial k_1 ^\alpha \partial k_2 ^\beta}
p\!\!/ p^\rho p^\sigma\right]_{k_1=x_1p, k_2=x_2p} \right.\nonumber\\
&&\hspace{0.5in} \times
  \langle A \mid \bar{\psi}(0)n\!\!\!/ \vec \partial_\bot A_+ (y_1^-)
\cdot\vec\partial_\bot A_+(y^-_2)\psi(y^-) \mid A\rangle
 \nonumber\\
&+& d^{\alpha\sigma}
{\rm Tr}\left[\frac{\partial\hat{H}_{\rho \sigma}(k,x_1p,k_2)}
{\partial k_2^\alpha}
\frac{p\!\!/p^\rho}{x_1}\right]_{k_2=x_2p} \nonumber\\
&&\hspace{0.5in} \times
  \langle A \mid \bar{\psi}(0)n\!\!\!/ \partial _+\vec A_\bot(y_1^-)
\cdot\vec\partial_\bot A_+(y^-_2)\psi(y^-)
 \mid A\rangle
  \nonumber\\
& + & d^{\alpha\rho}
{\rm Tr}\left[\frac{\partial\hat{H}_{\rho \sigma}(k,k_1,x_2p)}
 {\partial k_1^\alpha}
\frac{p\!\!/p^\sigma}{x_2}\right]_{ k_1=x_1p} \nonumber\\
&&\hspace{0.5in} \times
  \langle A \mid \bar{\psi}(0)n\!\!\!/ \vec\partial_\bot A_+(y^-_1)
\cdot \partial_+\vec A_\bot(y_2^-)\psi(y^-) \mid A\rangle
 \nonumber\\
 &+ &d^{\rho\sigma} {\rm Tr}\left[\hat{H}_{\rho \sigma}(k,x_1p,x_2p)
\frac{p\!\!/}{x_1x_2}\right] \nonumber\\
&&\hspace{0.5in} \times
 \langle A \mid \bar{\psi}(0)n\!\!\!/ \partial_+\vec A_\bot (y^-_1)
\cdot\partial _+\vec A_\bot(y_2^-)\psi(y^-) \mid A\rangle \  \}.
\label{fact1}
\end{eqnarray}
Using the following identities which we will prove in the next
section,
\begin{eqnarray}
 & &d^{\alpha\beta}
\frac{\partial ^2}{\partial k_1 ^\alpha \partial k_2 ^\beta}
\left. \hat{H}_{\rho \sigma}(k,k_1,k_2)p^\rho p^\sigma
\right|_{k_1=x_1p, k_2=x_2p}
\nonumber\\
&&\hspace{1.2in}
=d^{\rho\sigma} \hat{H}_{\rho\sigma}(k,x_1p,x_2p) \frac{1}{x_1x_2}
\nonumber \\
&&\hspace{1.2in}
 = -d^{\alpha\sigma} \frac{\partial}{\partial k_2^\alpha}
\hat{H}_{\rho \sigma}(k,x_1p,k_2) \mid _{ k_2=x_2p} p^\rho \frac{1}{x_1}
\nonumber \\
&&\hspace{1.2in}
=-d^{\alpha\rho} \frac{\partial}{\partial k_1^\alpha}
\hat{H}_{\rho \sigma}(k,k_1,x_2p)\mid _{ k_1=x_1p}p^\sigma
\frac{1}{x_2} \label{idd2}
\end{eqnarray}
one can reorganize the hadronic tensor in the covariant gauge as
\begin{eqnarray}
W^{(2)}&=& \frac{1}{8} \int
 \frac{dy^-}{2\pi}\frac{dy^-_1}{2\pi}\frac{dy^-_2}{2\pi}\int dx_1dx_2dx
e^{ix_1p^+y_1^- + ix_2p^+y_2^- + ixp^+y^-}
\nonumber\\
& \times &
d^{\alpha\beta}{\rm Tr}\left[\frac{\partial ^2\hat{H}^{\rho\sigma}(k,k_1,k_2)}
{\partial k_1 ^\alpha \partial k_2 ^\beta} p\!\!/
p_\rho p_\sigma\right]_{k_1=x_1p, k_2=x_2p}
\langle A \mid \bar{\psi}(0)n\!\!\!/ \nonumber\\
&\times& \left \{ \vec\partial_\bot A_+
(y_1^-)\cdot\vec\partial_\bot A_+(y^-_2) - \partial_+\vec A_\bot(y_1^-)
\cdot\vec\partial_\bot A_+(y^-_2) \right. \nonumber\\
&& \left.  -\vec\partial_\bot A_+(y^-_1)\cdot\partial _+\vec A_\bot(y_2^-) +
\partial_+\vec A_\bot (y^-_1)\cdot\partial_+\vec A_\bot(y_2^-) \right \} \psi(y^-)
\mid A\rangle \nonumber \\
&=& \int
 \frac{dy^-}{2\pi}\frac{dy^-_1}{2\pi}\frac{dy^-_2}{2\pi}\int dx_1dx_2dx
e^{ix_1p^+y_1^- + ix_2p^+y_2^- + ixp^+y^-}\nonumber\\
& & \frac{d^{\alpha\beta}}{4}
 {\rm Tr}\left[\frac{\partial ^2 \hat{H}^{\rho \sigma}(k,k_1,k_2)}
 {\partial k_1 ^\alpha \partial k_2 ^\beta}
p\!\!/p_\rho p_\sigma\right]_{k_1=x_1p, k_2=x_2p}
\nonumber\\
&& \hspace{1.0in}
\times \langle A \mid \bar{\psi}(0)
\frac{n\!\!\!/}{2}\vec F_{+\bot}(y^-_1)\cdot\vec F_{+\bot}(y^-_2)
\psi(y^-)\mid A\rangle \, ,
\label{eq:cow}
\end{eqnarray}
where we have neglected higher order corrections to the definition
of the gluon field strength $F_{+\bot}$ in the covariant gauge.
In the original derivation of the twist expanion framework \cite{LQS},
LQS neglected terms that are proportional to the transverse
gluon fields in the covariant gauge in Eq.~(\ref{fact1}), arguing
that these contributions are suppressed by $1/p^+$. As we have
shown in the above, the corresponding hard partonic parts of
all these contributions are equivalent, allowing us to
combine all terms. The final result is proportional to the
correlation of total field strength with one common hard partonic
parts. Since the gluon field strength correlator is now in a
gauge invariant form (with gauge links from other soft gluon
interaction), the hard partonic parts should also be gauge
invariant and can be calculated in any gauge.

Comparing Eqs.~(\ref{eq:lgw}) and (\ref{eq:cow}), the leading
nuclear enhanced twist-4 contributions to the hadronic tensor
of DIS off a large nucleus in both light-cone and covariant gauge
have the same non-perturbative matrix elements for two-parton
correlation functions. Since these matrix elements are gauge
invariant, the corresponding
hard parts should be equivalent and also gauge invariant, therefore
can be calculated in any given gauge. The equivalence of the
corresponding hard parts can be proved in all orders as given by
the identity in Eq.~(\ref{idd2}).

%%%%%%%%%%%%%%%%%%%%%%%%%%%%%%%%%%%%%%%%%%%%%%%%%%%%%%%%%%%%%%%%%%%%%%%%%%%%%%%%%%%%%%%%%%%%%%%%%%%%%%$$$$$$%%%%%%%%%%%%

\section {Equivalence of hard parts}

In this section we will prove the identities in Eq.~(\ref{idd2}).
The technique of Ward identity has been used frequently in the proof of
factorization in pQCD hard processes in which longitudinal gluons can
be factorized from the hard part and give rise to eikonal lines in hard
scattering \cite{collins}. This method is also the basic ingredient
in our proof of the identities that are used to prove the equivalence of
the hard parts in the leading twist-4 contributions to DIS off a large
nucleus in the light-cone and covariant gauge.

\begin{figure}
\begin{center}
\includegraphics[width=4.0in]{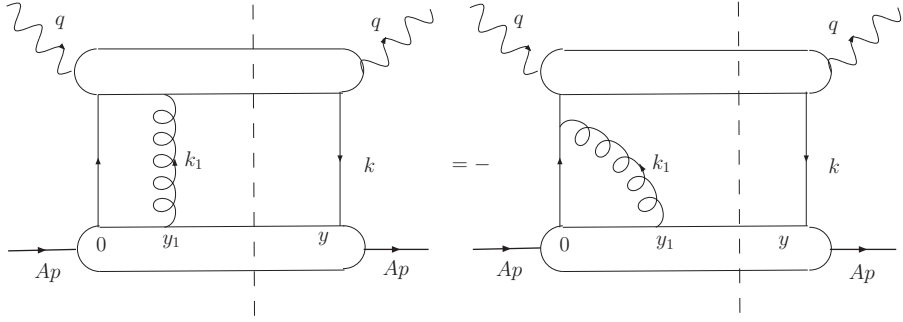}
\end{center}
\caption{Diagrammtic Ward identity(dash line represent the cut)}
\label{fig2}
\end{figure}

We first consider a right-cut diagram with a single external gluon
attached to the hard part in Fig.~\ref{fig2}. The gluon momentum is
${k^{\rho}_1}=x_1p^{\rho}+k^{\rho}_{1\bot}$. We again neglect the
minus component of the gluon momentum which only contributes to
higher twist terms. Assuming the total amplitude $\hat{\cal M}^\rho$
that includes all possible gluon attachment to the hard part, gauge
invariance, $\hat{\cal M}_\rho k_1^\rho=0$, leads to the following
Ward identity,
\begin{equation}
\hat{H_\rho}(k,k_1) k_1^\rho u(k-k_1)=-\hat{H}(k)
\frac{i}{k\!\!\!/} ik \!\!\!/ _1 u(k-k_1),
\end{equation}
as shown diagrammatically in Fig.~\ref{fig2},
where $u(k)$ is the quark spinor, $\hat{H_\rho}(k,k_1)$ represents
the hard part with one gluon attached after the photon coupling
and $\hat{H}(k)$ represents the hard part without any gluon
attachment. Using
equation of motion $(k\!\!\!/-k \!\!\!/ _1)u(k-k_1)=0$, we get,
\begin{equation}
\hat{H_\rho}(k,k_1) k_1^\rho u(k-k_1)
=\hat{H}(k) u(k-k_1)
\end{equation}
Truncating the quark spinor which sits inside the non-perturbative
parton matrix elements, we have the following general identity which
is also valid for the diagram with the off-shell initial-state quarks,
\begin{equation}
\hat{H_\rho}(k,k_1) k_1^\rho=\hat{H}(k)
\label{iden1}
\end{equation}
Making a collinear expansion in $k_1^\rho=x_1p^{\rho}+k^{\rho}_{1\bot}$
of the left side,
\begin{eqnarray}
\hat{H_\rho}(k,k_1) k_1^\rho=\hat{H_\rho}(k,x_1p)x_1p^{\rho}
&+& \hat{H_\rho}(k,x_1p)k_{1\bot}^\rho
\nonumber \\
&+&
\left. \frac{\partial{\hat{H_\rho}(k,k_1)}} {\partial{k_1^\alpha}}
\right|_{k_1=x_1p} k_{1\bot}^\alpha x_1p^\rho+....
\label{iden2}
\end{eqnarray}
and using Eq.~(\ref{iden1}) for both $k_1$ and $x_1p$,
$\hat{H_\rho}(k,x_1p)x_1p^\rho=\hat{H}(k)=\hat{H_\rho}(k,k_1)k_1^\rho$,
we obtain the following identity,
\begin{equation}
\left. \frac{\partial{\hat{H_\rho}(k,k_1)}} {\partial{k_1^\alpha}}
\right|_{k_1=x_1p} k_{1\bot}^\alpha x_1p^\rho=
-\hat{H_\rho}(k,x_1p)k^{\rho}_{1\bot}
\end{equation}
Note that authors of the Ref~\cite{Eguchi:2006mc} have recently derived
the above identity case by case.
%%%%%%%%%%%%%%%%%%%%%%%%%%%%%%%%%%%%%%%%%%%%%%%%%%%%%%%%%%%%%%%%%%%%%%%%%%%%%%%%%%%%%%%%%%%%%%%%%%%%%%%%%%%%%%%%%%%%%

%\begin{figure}[htbp]
%\begin{center}
%\includegraphics[width=2.5in,height=1.8in]{fig2.eps}
%\end{center}
%\caption{ Two gluons attachment}
%\label{fig3}
%\end{figure}

For diagrams with two attached gluons that contribute to twist-4 terms,
one can derive two identities  similar to Eq.~(\ref{iden1}) by
contracting with the momentum of gluon 1 and  gluon 2, $ k_1 $, $ k_2$
and then make collinear expansion in them respectively,
\begin{eqnarray}
\left. \frac{\partial{\hat{H}_{\rho\sigma}(k,k_1,k_2)}}
{\partial{k_1^\alpha}}\right|_{k_1=x_1p}\!\!\!\! k_{1\bot}^\alpha x_1p^\rho
&=&-\hat{H}_{\rho\sigma}(k,x_1p,k_2)k^{\rho}_{1\bot}
\label{iden21} \\
\left. \frac{\partial{\hat{H}_{\rho\sigma}(k,k_1,k_2)}}
{\partial{k_2^\beta}}\right|_{k_2=x_2p}\!\!\!\! k_{2\bot}^\beta x_2p^\sigma
&=&-\hat{H}_{\rho\sigma}(k,k_1,x_2p)k^{\sigma}_{2\bot}
\label{iden22}
\end{eqnarray}
Making again collinear expansion in $k_{2\bot}$ on both sides
of Eq.~(\ref{iden21}), multiplying both sides with $x_2p^\sigma$
and using Eq.~(\ref{iden22}) with $k_1=x_1p$, one has
\begin{eqnarray}
\left. \frac{\partial^2{\hat{H}_{\rho\sigma}(k,k_1,k_2)}}
{\partial{k_1^\alpha}\partial{k_2^\beta}} \right|_{k_1=x_1p,k_2=x_2p}&&
\hspace{-0.3in}k_{1\bot}^\alpha k_{2\bot}^\beta x_1p^\rho x_2p^\sigma \nonumber \\
&&\hspace{-0.3in}=\left. -\frac{\partial \hat{H}_{\rho\sigma}(k,x_1p,k_2)}
{\partial k_2^\beta}
\right|_{k_2=x_2p}\!\!\!\! k_{2\bot}^\beta k^{\rho}_{1\bot}
x_2p^\sigma
\label{iden31} \nonumber \\
&&=\hat{H}_{\rho\sigma}(k,x_1p,x_2p)
k_{1\bot}^\rho k^{\sigma}_{2\bot}
\end{eqnarray}
Similarly making collinear expansion in $k_{1\bot}$ on both sides
of Eq.~(\ref{iden22}), multiplying both sides with $x_1p^\rho$
and using Eq.~(\ref{iden21}) with $k_2=x_2p$ gives,
\begin{eqnarray}
\left. \frac{\partial^2{\hat{H}_{\rho\sigma}(k,k_1,k_2)}}
{\partial{k_1^\alpha}\partial{k_2^\beta}}\right|_{k_1=x_1p,k_2=x_2p}&&
\hspace{-0.3in}
k_{1\bot}^\alpha k_{2\bot}^\beta x_1p^\rho x_2p^\sigma \nonumber \\
&&=\left. -\frac{\partial \hat{H}_{\rho\sigma}(k,x_1p,k_2)}
{\partial k_1^\alpha}\right|_{k_1=x_1p}\!\!\!\!
k_{1\bot}^\alpha k^{\sigma}_{2\bot} x_1p^\rho \nonumber \\
&&=\hat{H}_{\rho\sigma}(k,x_1p,x_2p)k_{1\bot}^\rho k^{\sigma}_{2\bot}
\label{iden32}
\end{eqnarray}
We can further integrate the above two sets of equations over the
azimuthal angle of the gluons' transverse momentum, divide both sides
by $x_1x_2$ and obtain finally the identities in Eqs.(\ref{idd2}),
\begin{eqnarray}
d^{\alpha\beta}\left. \frac{\partial^2\hat{H}^{\rho\sigma}(k,k_1,k_2)}
{\partial k_1 ^\alpha \partial k_2 ^\beta} p_\rho p_\sigma
\right|_{k_1=x_1p, k_2=x_2p}
&&\hspace{-0.3in}=d^{\rho\sigma} \hat{H}^{\rho\sigma}(k,x_1p,x_2p) \frac{1}{x_1x_2}\\
&&\hspace{-0.7in}
=-d^{\alpha\sigma} \left. \frac{\partial\hat{H}^{\rho\sigma}(k,x_1p,k_2)}
{\partial k_2^\alpha}\frac{p_\rho}{x_1}\right| _{k_2=x_2p} \\
&&\hspace{-0.7in}
=-d^{\alpha\rho}\left. \frac{\partial\hat{H}^{\rho \sigma}(k,k_1,x_2p)}
{\partial k_1^\alpha}\frac{p_\sigma}{x_2} \right|_{k_1=x_1p}\, .
\end{eqnarray}
%%%%%%%%%%%%%%%%%%%%%%%%%%%%%%%%%%%%%%%%%%%%%%%%%%%%%%%%%%%%%%%%%%%%%%%%%%%%%%%%%%%%%%%%%%%%%%%%%%%%%%%%%%%%%%%%%%%%

\section{DIS in leading order}

The general proof of the identities for the hard parts in the previous
section is valid for all orders in $\alpha_{\rm s}$.
In this and next section, we will calculate nuclear modification to
the semi-inclusive cross section of DIS off a large nucleus up to
$\cal{O}$$(\alpha_s)$ as an explicit example to demonstrate the
gauge invariance.

\begin{figure}[htbp]
\begin{center}
\includegraphics[width=3.5in]{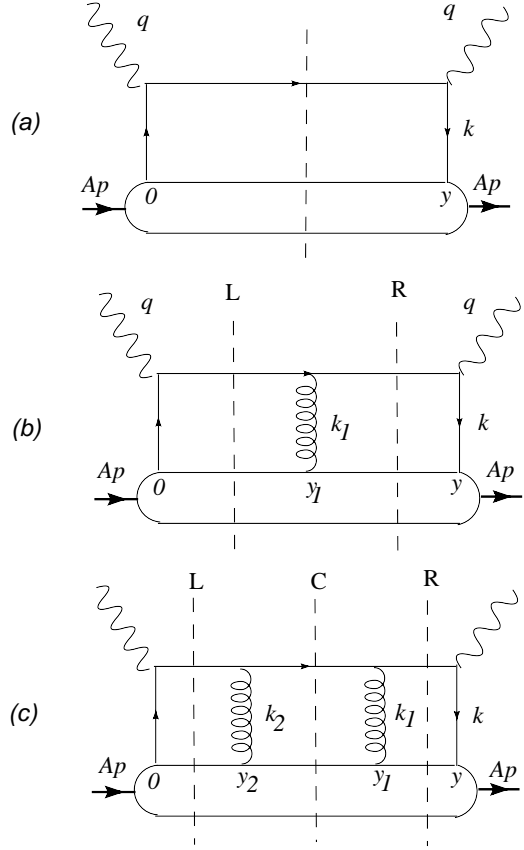}
\end{center}
\caption{DIS in lowest order with zero, one and two gluon exchange between the final
quark and the target remnant}
\label{fig3}
\end{figure}

The semi-inclusive cross section off DIS in Eq.~(\ref{sidis}) at the lowest
order in $\alpha_s$ has been calculated in Ref.~\cite{Liang:2006wp} up to two
gluon exchanges as shown in Fig.~\ref{fig3}.
The semi-inclusive hadronic tensor can be expanded in terms
of the number of physical gluon exchanges \cite{Liang:2006wp}
$W_{\mu\nu}(q,p,\ell)=\sum_i W_{\mu\nu}^{(i)}(q,p,\ell)$,
\begin{eqnarray}
\label{eq:Wsi1}
\frac{d^2W_{\mu\nu}^{(0)}}{d^2\ell_\perp}
&=&\frac{1}{2\pi}
\int\frac{d^4k}{(2\pi)^4}\delta^{(2)}(\vec\ell_\perp -\vec k_\perp)
{\rm Tr}[\hat H_{\mu\nu}^{(0)}(x)\ \hat \Phi^{(0)}(k)];\\
%%%%%%%
\label{eq:Wsi2}
\frac{d^2W_{\mu\nu}^{(1)}}{d^2\ell_\perp}
&=&\frac{1}{2\pi}
\int\frac{d^4k}{(2\pi)^4}\frac{d^4k_1}{(2\pi)^4}
\sum_{c=L,R} \delta^{(2)}(\vec\ell_\perp -\vec k_{c\perp})
\nonumber \\
&& \times{\rm Tr}[\hat H_{\mu\nu}^{(1,c)\rho}(x,x_1) \omega_\rho^{\ \rho'}
\hat \Phi^{(1)}_{\rho'}(k_1,k)];\\
%%%%%%%%
\label{eq:Wsi3}
\frac{d^2W_{\mu\nu}^{(2)}}{d^2\ell_\perp}
&=&\frac{1}{2\pi}
\int\frac{d^4k_1}{(2\pi)^4}\frac{d^4k_2}{(2\pi)^4}\frac{d^4k}{(2\pi)^4}
\sum_{c=L,R,C}\delta^{(2)}(\vec\ell_\perp -\vec k_{c\perp})
\nonumber \\
&&\times {\rm Tr}[\hat H_{\mu\nu}^{(2,c)\rho\sigma}(x,x_1,x_2)
\omega_\rho^{\ \rho'}\omega_\sigma^{\ \sigma'}
\hat\Phi^{(2)}_{\rho'\sigma'}(k_1,k_2,k)],
\end{eqnarray}
where the summation is over different cut diagrams with given number
of gluon exchanges, $\hat H_{\mu\nu}^{(0)}$, $\hat H_{\mu\nu}^{(1,c)\rho}$ and
$\hat H_{\mu\nu}^{(2,c)\rho\sigma}$ are the hard parts of the
corresponding cut diagrams (see Ref.~\cite{Liang:2006wp} for details).
For single gluon exchange, $\vec k_{R\perp}=\vec k_\perp$, 
$\vec k_{L\perp}=\vec k_\perp+ \vec k_{1\perp}$
For two gluon exchange diagram, $\vec k_{R\perp}=\vec k_\perp$, 
$\vec k_{C\perp}=\vec k_\perp+\vec k_{1\perp}$
and $\vec k_{L\perp}=\vec k_\perp+\vec k_{1\perp}+ \vec k_{2\perp}$.
The projection operator is defined such that
$\omega_\rho^\sigma k_{i\sigma}=(k_i-x_ip)_\rho$. The above expression
for semi-inclusive cross section is obtained through collinear
expansion in $(k_i-x_ip)$ of the partonic hard parts and the decomposition
of the gauge field $A_\rho = p_\rho A^+/p^+ + \omega_\rho^\sigma A_\sigma$.
Generalized Ward identities are used to relate the derivatives of
the partonic hard parts to the hard parts with extra longitudinal
gluon attachments. Consequently, the unintegrated parton matrix elements contain
contributions from all Feynman diagrams with different number of
unphysical gluon exchanges between the propagating quark and the
nucleus. They can be defined in a gauge invariant way as,
\begin{eqnarray}
\label{eq:Phiun1}
\hat\Phi^{(0)}(k)&=&\int d^4ye^{ik\cdot y}
\langle A |\bar\psi(0){\cal L}(0,y) \psi(y)|A \rangle, \\
%%%%%%
\label{eq:Phiun2}
\hat\Phi^{(1)}_\rho(k_1,k)
&=&\int d^4yd^4y_1e^{ik\cdot y+ik_1\cdot y_1}
\langle A |\bar\psi(0) {\cal L}(0,y_1)D_\rho(y_1){\cal L}(y_1,y)\psi(y)| A \rangle,
\nonumber \\
\\
%%%%%%
\label{eq:Phiun3}
\hat\Phi^{(2)}_{\rho\sigma}(k_1,k_2,k)&=&\int d^4yd^4y_1d^4y_2
e^{ik\cdot y+ik_1\cdot y_1+ik_2\cdot y_2} \nonumber \\
&\times& \langle A |\bar\psi(0){\cal L}(0,y_2)
D_\rho(y_2) {\cal L}(y_2,y_1)D_\sigma(y_1){\cal L}(y_1,y)\psi(y)| A \rangle,
\nonumber \\
\end{eqnarray}
where $D_\rho(y)=\partial_\rho + ig A_\rho(y)$ is the covariant
derivative and
\begin{eqnarray}
{\cal L}(y_2,y_1)&\equiv&
{\cal L}^\dagger_\parallel(-\infty,y_2^-;\vec y_{2\perp})
{\cal L}^\dagger_\perp(-\infty;\vec y_{2\perp},\vec y_{1\perp})
{\cal L}_\parallel(-\infty,y^-_1;\vec y_{1\perp})
\end{eqnarray}
is the complete gauge link that contains both the
transverse \cite{Belitsky:2002sm}
\begin{equation}
{\cal L}_\perp(-\infty;\vec y_{2\perp},\vec y_{1\perp})
\equiv P\exp\left[-ig\int_{\vec y_{2\perp}}^{\vec y_{1\perp}}
d\vec\xi_\perp\cdot
\vec A_\perp(-\infty,\vec\xi_\perp)\right]
\end{equation}
and longitudinal gauge link
\begin{equation}
{\cal L}_\parallel (-\infty,y^-;\vec y_\perp)\equiv
P\exp\left[-ig\int^{-\infty}_{y^-} d\xi^- A_+(\xi^-,\vec y_\perp)\right].
\end{equation}
The exchange of unphysical gluon between the propagating quark and the
nucleus therefore leads to the gauge links in the parton matrix
elements while physical gluons lead to the higher-twist contributions
to the DIS cross section which are characterized by the covariant
derivatives $D_\rho$ in the higher-twist parton
matrix elements. The spatial derivative in $D_\rho$ comes from
the collinear expansion of the partonic hard parts while the
transverse gluon field in $D_\rho$ corresponds to interaction
between the propagating quark and a physical gluon from the nucleus.

We refer the above organization of semi-inclusive DIS cross
section within the collinear expansion as a generalized
twist-expansion in which each contribution involves a parton
matrix element with a given number of covariant derivatives.
These parton matrix elements correspond to unintegrated parton
distributions. Note that after integration (over both $k^-$ and
$\vec k_\perp$), each of these unintegrated parton distributions
will give rise to a mixture of collinear (or integrated) parton
distributions or parton matrix elements with different dimensions,
because of the off-shellness and transverse momentum carried by each parton.

The nuclear enhanced twist-four contributions to the semi-inclusive
cross section at the leading order is contained in Eq.~(\ref{eq:Wsi3})
with the corresponding projected parton matrix elements
$\omega_{\rho}^{\rho^\prime} \omega_{\sigma}^{\sigma^\prime}
\hat\Phi^{(2)}_{\rho^\prime\sigma^\prime}(k_1,k_2,k)$ in Eq.~(\ref{eq:Phiun3})
which is apparently gauge invariant. Note that the leading
components of the projected covariant derivative
$\omega_{\rho}^{\rho^\prime} D_{\rho^\prime}$ are
the transverse ones, $\vec D_\perp=\vec \partial_\perp +ig\vec A_\perp$.
Using the following identity \cite{Liang:2008vz},
\begin{eqnarray}
i\vec\partial_{y_\perp}{\cal L}(y_1,y)
&=&{\cal L}(y_1,y)\left[ i\vec D_\perp(y^-,\vec y_\perp) \right.
\nonumber \\
&+& \left. g\int_{-\infty}^{y^-}d\xi^-
{\cal L}^\dagger_\parallel(\xi^-,y^-;\vec y_\perp)\vec F_{+\perp}(\xi^-,\vec y_\perp)
{\cal L}_\parallel(\xi^-,y^-;\vec y_\perp)\right],
\label{iden}
\end{eqnarray}
it is easy to see that the parton matrix element
$\omega_{\rho}^{\rho^\prime} \omega_{\sigma}^{\sigma^\prime}
\hat\Phi^{(2)}_{\rho^\prime\sigma^\prime}(k_1,k_2,k)$ in
Eq.~(\ref{eq:Phiun3}) contains exactly the same quark-gluon
correlation distribution in Eqs.~(\ref{eq:lgw}) and (\ref{eq:cow}).
Note that the hard part
in Eq.~(\ref{eq:Wsi3}) is projected to the transverse polarization
of the gluon exchange and therefore corresponds to interaction with
collinear physical gluons in the light-cone gauge. In the 
derivation \cite{Liang:2006wp} of this final form, Ward indentities are 
also used to relate it to the derivative of the hard parts for quark and 
longitudinal gluon interaction in the covariant gauge.

After integration over the transverse momentum of the quark, one will be able
to recover the nuclear enhanced twist-four contributions 
to the inclusive DIS cross section \cite{Ellis:1982wd,Qiu:1988dn}, which
should be power-suppressed by $1/Q^2$. 
In the calculation of the above twist-four contribution to the inclusive 
DIS cross section, one can similarly
relate the transverse gauge field $\vec A_\perp$ to the field
strength $\vec F_{+\perp}$ via partial integration. However,
one should include the gluonic poles \cite{Boer:1997bw} in
\begin{equation}
\int dy^- e^{ixp^+y^-}A_\perp(y^-)
=\int dy^- e^{ixp^+y^-}\frac{1}{x\pm i\epsilon}F_{+\perp}(y^-),
\end{equation}
depending on the boundary condition for $A_\perp(\pm\infty)$ which
do not vanish simultaneously in the light-cone gauge \cite{Ji:2002aa}.
We have chosen $\vec A_\perp(\infty)=0$ in this paper.
Such gluonic poles have non-vanishing contributions to
higher-twist inclusive cross sections of DIS which are suppressed by $1/Q^2$.
Without inclusion of these gluonic pole contributions, one
could be misled to the conclusion \cite{fries} that interaction
with physical transverse gluons in the light-cone gauge could lead to leading
twist transverse momentum broadening of a propagating quark, which
does not contribute to higher-twist ($1/Q^2$ suppressed) inclusive
cross section. As we will show in the next section, such
transverse momentum broadening comes from interaction with
unphysical gluons (either longitudinal gluon $A_+$ in covariant
gauge or pure transverse gauge field $\vec A_\perp(-\infty)$ at the
light-cone infinity in the light-cone gauge).

\section {Nuclear transverse momentum broadening}

Even though the exchange of unphysical gluons do not contribute
to the hard parts of higher-twist DIS cross sections, they do lead to
important gauge links in parton distributions and the transverse
momentum broadening of the propagating quark inside a nucleus
at the leading twist (no $1/Q^2$ suppression). Exchange of physical 
gluons (e.g. transverse gluons) only leads to higher-twist inclusive 
cross section and transverse momentum broadening
which is suppressed by factors of $1/Q^2$ as compared to the
leading twist broadening. This has been a source of confusion
in previous studies.

Consider the semi-inclusive hadronic tensor with no physical gluon
exchange in Eq.~(\ref{eq:Wsi1}). The collinear component of the
parton matrix element
\begin{equation}
\hat\Phi^{(0)}(k)=\frac{1}{2}p\!\!\!/ \hat f_A^q(k)+
\frac{1}{2}(k\!\!\!/ -xp\!\!\!/)\hat f_{\perp A}^q(k)
\end{equation}
will give rise to the leading twist contribution to the semi-inclusive
hadronic tensor
\begin{equation}
\frac{d^2W_{\mu\nu}^{(0)}}{d^2\ell_\perp}
=\frac{1}{2\pi} \int dx d^2k_\perp\delta^{(2)}(\vec\ell_\perp -\vec k_\perp)
{\rm Tr}[\hat H_{\mu\nu}^{(0)}(x)\frac{p\!\!\!/}{2} ]f_A^q(x,\vec k_\perp);\\
%%%%%%%
\end{equation}
and the transverse momentum dependent quark distribution function
\begin{eqnarray}
f^q_A(x,\vec k_\perp) &=& \int dk^- \hat f_A^q(k)
=\int \frac{dy^-}{2\pi} \frac{d^2y_\perp}{(2\pi)^2}
e^{ixp^+y^- -i\vec k_\perp\cdot \vec y_\perp}
\nonumber \\
&\times& \langle A \mid \bar\psi(0,\vec 0_\perp)\frac{\gamma^+}{2}
{\cal L}(0,y) \psi(y^-,\vec y_\perp)
\mid A \rangle .
\label{tmd0}
\end{eqnarray}
The corresponding integrated quark distribution function is
\begin{eqnarray}
f_A^q(x)&=&\int d^2k_\perp f_A^q(x,\vec k_\perp) \nonumber \\
&=&\int \frac{dy^-}{2\pi} e^{ixp^+y^-}
\langle A \mid \bar\psi(0,\vec 0_\perp)\frac{\gamma^+}{2}
{\cal L}_{\parallel}(0,y^-) \psi(y^-,\vec 0_\perp)\mid A \rangle .
\end{eqnarray}
One can make a Taylor expansion of the gauge link and the quark field
in the transverse coordinate $\vec y_\perp$ and then complete
integration over $\vec y_\perp$,
\begin{eqnarray}
f^q_A(x,\vec k_\perp)&=&\int \frac{dy^-}{2\pi}
e^{ixp^+y^-} \langle A \mid \bar\psi(0,\vec 0_\perp)\frac{\gamma^+}{2}
\nonumber \\
&&\hspace{0.5in}\times\left[ e^{i\vec\partial_{y_\perp}\cdot\vec\partial_{k_\perp}}
{\cal L}(0,y) \psi(y^-,\vec y_\perp)\right]_{y_\perp=0}
\mid A \rangle \delta^{(2)}(\vec k_\perp).
\label{tmdf1}
\end{eqnarray}
Using the identity in Eq.~(\ref{iden}), one can express the above
transverse momentum dependent quark distribution in terms of collinear
higher-twist quark and gluon correlation matrix
elements \cite{Liang:2008vz}.

For the purpose of discussion in this paper,
we consider first the quadratic term in the
expansion of the gauge link in the covariant gauge,
\begin{eqnarray}
{\cal L}^{(2)}(0,y)&=&\int dy^-_1 dy^-_2
\left[-g^2
A_+(y^-_2,\vec y_\perp)
A_+(y^-_1,\vec y_\perp) \theta(y^- -y_2^-)\theta(y^-_2 - y_1^-)\right.
\nonumber \\
&+&g^2 A_+(y^-_2,\vec 0_\perp)
A_+(y^-_1,\vec y_\perp) \theta(-y_2^-)\theta(y^- - y_1^-)
\nonumber \\
&-&\left. g^2  A_+(y^-_2,\vec 0_\perp)
A_+(y^-_1,\vec 0_\perp) \theta(-y_1^-)\theta(y^-_1 - y_2^-)
\right].
\end{eqnarray}
The three terms in the above expansion correspond to the left,
central and right cut diagrams
in Fig.~\ref{fig3}(c). One can further expand the derivative operator
$\exp(i \vec\partial_{y_\perp}\cdot\vec\partial_{k_\perp})$ to
the quadratic term.
It is easy to note that only the left-cut contribution
has a term $F_{+\perp}(y_2^-)F_{+\perp}(y^-_1)\approx
\partial_\perp A_+(y_2^-,0_\perp)\partial_\perp A_+(y_1^-,0_\perp)$
from the quadratic derivative
 $(\vec\partial_{y_\perp}\cdot\vec\partial_{k_\perp})^2$.
As we will explain later this is the leading term that has a nuclear
enhancement.

One can also express the above contributions
explicitly in terms of the transverse momentum of each gluonic field
as denoted in Fig.~\ref{fig3}(c),
\begin{eqnarray}
&&{\cal L}^{(2)}(0,y)\
%psi(y^-,\vec y_\perp)
%e^{-i\vec k_\perp\cdot\vec y_\perp}
=\int d^2k_{1\perp} d^2k_{2\perp}
\int \frac{d^2y_{1\perp}}{(2\pi)^2}\frac{d^2y_{2\perp}}{(2\pi)^2}
e^{-i\vec k_{1\perp}\cdot\vec y_{1\perp}
-i\vec k_{2\perp}\cdot\vec y_{2\perp}} \nonumber \\
&&\hspace{0.5in} \times
D (\vec k_{1\perp},\vec k_{2\perp},\vec k_\perp,\vec\ell_\perp)
A_+(y^-_2,\vec y_{2\perp})A_+(y^-_1,\vec y_{1\perp})
%\psi(y^-,\vec y_\perp),
\nonumber \\
&&D \equiv
\int dy^-_1 dy^-_2 \left[ -g^2 \theta(y^- -y_2^-)\theta(y^-_2 - y_1^-)
\delta^{(2)}(\vec\ell_\perp -\vec k_\perp - \vec k_{1\perp} - \vec k_{2\perp})
\right.
\nonumber \\
&&\hspace{0.5in} +g^2 \theta(-y_2^-)\theta(y^- - y_1^-)
\delta^{(2)}(\vec\ell_\perp -\vec k_\perp - \vec k_{1\perp})
\nonumber \\
&&\hspace{0.5in}\left. -g^2 \theta(-y_1^-)\theta(y^-_1 - y_2^-)
\delta^{(2)}(\vec\ell_\perp -\vec k_\perp)\right].
\end{eqnarray}
One can expand the above $\delta$-functions in the transverse
momenta of the quark and gluon fields $\vec k_\perp$, $\vec k_{1\perp}$
and $\vec k_{2\perp}$. Contribution to
 $\partial_\perp A_+(y_2^-,0_\perp)\partial_\perp A_+(y_1^-,0_\perp)$
again comes only from the first term which corresponds to the
left-cut diagram in Fig.~\ref{fig3}(c). Since the dependence of
the partonic hard part $D$ on the transverse momentum of the quark field
only leads to the higher-twist matrix elements that are not nuclear
enhanced, one can set $k_\perp=0$ in the hard part. By momentum
conservation $\vec k_{1\perp}=-\vec k_{2\perp}$. In this case,
only the second term from the central-cut diagram contribute to
the parton matrix
$\partial_\perp A_+(y_2^-,0_\perp)\partial_\perp A_+(y_1^-,0_\perp)$
for the leading nuclear broadening.

One can make a similar analysis of the transverse momentum
distribution in the light-cone gauge. The leading twist
contribution comes from the transverse gauge link which is
determined by the boundary condition of the transverse gauge field
$A_\perp(-\infty, \vec y_\perp)$ \cite{Liang:2008vz}.
As we have discussed before, quark interaction with physical
transverse gauge field will lead to higher-twist inclusive 
cross section and the transverse momentum broadening which 
is power suppressed by $1/Q^2$ as compared to the leading twist
contribution.

Taking the quadratic derivatives of the two-gluon contribution
[Fig.~\ref{fig3}(c)] to the gauge link in Eq.~(\ref{tmdf1}), one
can obtain the corresponding transverse momentum distribution,
\begin{eqnarray}
f_A^q(x,\vec \ell_\perp)
&\approx &f_A^q(x)\delta^{(2)}(\vec\ell_\perp)+
\frac{2\pi\alpha_s}{N_c}T_{qg}^{A}(x,0)
\frac{1}{4}\nabla^2_{\ell_\perp}\delta^{(2)}(\vec\ell_\perp)
\, ,
\label{eq:cov-w1}
\end{eqnarray}
where
\begin{eqnarray}
T^{A}_{qg}(x,x_1)&=&\int \frac{dy^-}{2 \pi}  dy^-_1 dy^-_2
e^{ixp^+y^- + ix_1p^+(y_1^- -y_2^-)}
\theta(-y_2^-)\theta(y^--y_1^-)
\nonumber\\
&&\frac{1}{2}\langle A \mid \bar{\psi}(0)
\gamma^+ \vec F_{+\perp}(y^-_2)\cdot \vec F_{+\perp}(y^-_1)
\psi(y^-)\mid A \rangle,
\label{tqg}
\end{eqnarray}
is the quark-gluon correlation function inside the nucleus.

The quark-gluon correlation function inside a nucleus
in Eq.~(\ref{tqg}) has two types of
contributions. Because a nucleus consists of nucleons which
are color singlet states, the quark-gluon pair could either come from
a single nucleon or from two separate nucleons. In the first case, all
four parton fields in the above correlation matrix element are confined
to the size of a nucleon $y^-, y_1^-, y_2^- \sim r_N$. On the other hand,
if quark and gluon fields are confined to two separate nucleons,
$y^-, |y_1^- -y_2^-| \sim r_N$, the overall position of the gluon
field will follow the second nucleon and are only confined to the size
of the nucleus $R_A$. Therefore, the quark-gluon correlation function in this
case will have a nuclear enhancement of the order $R_A/r_N\sim A^{1/3}$ as
compared to the first case where both quark and gluon fields are confined
to a single nucleon. For multiple scattering in a large nucleus, we will
only keep the second correlation with nuclear enhancement. If we further
neglect the correlation between different nucleons and assume the large
nucleus as a weakly bound and homogenous system of nucleons, the leading
contribution to the quark-gluon correlation function can be
approximated as \cite{owz,jorge}
\begin{eqnarray}
T_{qg}^{A}(x,x_1)&\approx& f^q_A(x) \int d\xi^-_N
\int \frac{d^3p_N}{(2\pi)^3 2p^+_N}
f_A(p_N,\xi_N)d\xi^- e^{ix_1p^+\xi^-} \nonumber \\
&\times&
\langle N \mid \vec F_{+\perp}(0)\cdot \vec F_{+\perp}(\xi^-)
\mid N \rangle \nonumber \\
&=&\pi f^q_A(x)\int d\xi^-_N \rho^A_N(\xi_N) x_1f^g_N(x_1),
\label{lgn-approx}
\end{eqnarray}
where $p_N^+=p^+$ is the longitudinal momentum per nucleon and
\begin{equation}
xf^g_N(x)=\int \frac{d\xi^-}{2\pi p^+} e^{ixp^+\xi^-}
\langle N \mid \vec F_{+\perp}(0)\cdot \vec F_{+\perp}(\xi^-)
\mid N \rangle ,
\label{gdis}
\end{equation}
is the gluon distribution function in a nucleon,
respectively. The spatial nucleon density inside the nucleus is
\[ \rho^A_N(\xi_N)=\int \frac{d^3p_N}{(2\pi)^3}f_A(p_N,\xi_N).\]
The integration over the spatial position of the nucleon
$\xi^-_N=(y_1^- +y_2^-)/2$ is limited to the size of the nucleus. The relative
coordinate of the two gluon fields is $\xi^-=y_1^- - y_2^-$.

With the above approximation, the transverse momentum broadening
squared of the propagating quark can be calculated as
\begin{equation}
\langle \Delta\ell_\perp^2\rangle \equiv
\frac{1}{f_A^q(x)}
\int d^2\ell_\perp \ell_\perp^2 f_A^q(x,\vec\ell_\perp)
=\int d\xi^-_N \hat q_F(\xi_N) ,
\label{ptbrod1}
\end{equation}
where the quark transport parameter
\begin{equation}
q_F(\xi_N)\equiv \frac{2\pi^2 \alpha_s}{N_c}
\rho^A_N(\xi_N) [xf_g^N(x)]_{x\approx 0}
\label{qhat1}
\end{equation}
can be interpreted as the broadening of
the mean transverse momentum squared per unit path length.
One can consider multiple gluon exchanges
the transverse distribution will become a Gaussian
form \cite{Liang:2008vz,fries,Majumder:2007hx} with the width
given by the averaged transverse momentum broadening.

In the above approximation of the twist-four quark-gluon matrix we have neglected
multiple-nucleon correlation in a large nucleus. Such an approximation is not
valid for small $x$ where quark-gluon and gluon-gluon fusion from different
nucleons become important and can lead to modification of the quark
distribution function and gluon saturation in a large
nucleus \cite{Gribov:1981ac,Mueller:1985wy}. One can take into account such effect
by using a nuclear modified quark distribution function $f_A^q(x_B)$ and
saturated gluon distribution function in the transport
parameter $\hat q_F$ which could lead a non-trivial nuclear and energy
dependence \cite{jorge}.

Note that the quark-gluon correlation function as defined in Eq.~(\ref{tqg})
and the gluon distribution function in Eq.(\ref{gdis}) are
not gauge invariant in covariant gauge. One needs to resum additional
number of collinear soft gluons on both side of the cut to produce
gauge links that will ensure the gauge invariance of the quark-gluon
correlation and the gluon distribution function. A general and gauge
invariant form of the transverse momentum broadening has been derived
in Ref.~\cite{Liang:2008vz}.

\section{Induced gluon radiation in light-cone gauge}

\begin{figure}
\begin{center}
\includegraphics[width=3.5in]{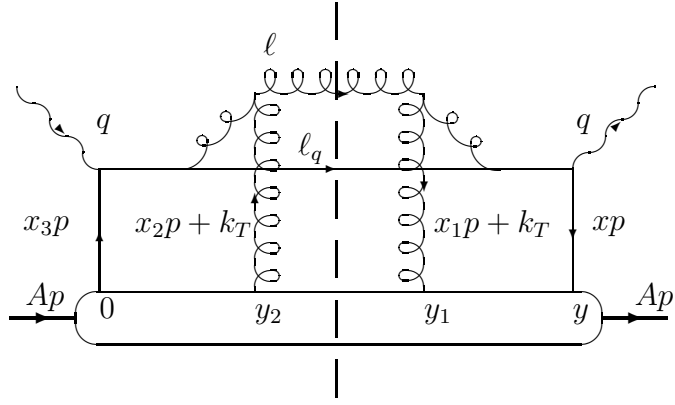}
\end{center}
\caption{Radiative correction to double hard scattering}
\label{fig4}
\end{figure}

As another example of the equivalence of the hard parts in double
parton scattering in light-cone and covariant gauge, we
consider the double hard quark-gluon scattering
in next leading order which has been calculated in covariant gauge
in helicity amplitude approximation in Ref.~\cite{gw-twist}. In the following
we will calculate the induced gluon spectra from such quark-gluon
scattering in light-cone gauge approach
within helicity amplitude approximation. Under such helicity amplitude
approximation, we can neglect momentum transfer to the quark, except in
its propagation direction and only consider its dominant minus component.
Since the initial gluon fields only have transverse components in the
light-cone gauge approach, their direct interaction with the quark will
produce a vertex in the form $n\!\!\!/ \epsilon\!\!/\!_\perp n\!\!\!/=0$
in the helicity amplitude approximation. Intuitively, this is
because a quark can not absorb the transversely polarized gluon due
to helicity conservation if we neglect the recoil induced by the
interaction. For this reason, we only need to consider the
quark-gluon rescattering with triple gluon vertex in the helicity
amplitude approximation as shown in Fig.~\ref{fig4}. Inclusion of
the quark recoil and other diagrams in light-cone gauge will lead
to power corrections in $\ell_\perp^2$ and $1-z$.

Contribution from the next-leading order double hard quark-gluon
scattering can be written as,
\begin{eqnarray}
\frac{dW_{\mu\nu}^{(2)}}{dz}&=&\sum_q
\int \frac{dy^-}{2 \pi} dy^-_1 dy^-_2 \int dx
\frac{dx_1}{2\pi}\frac{dx_2}{2\pi}
e^{ixp^+y^- +ix_1p^+y_1^- -ix_2p^+y_2^-}
  \nonumber\\
&&\hspace{-0.7in}\times
\frac{d_{\rho\sigma}}{2x_1x_2}
\widetilde{H}^{\rho\sigma}_{\mu\nu} (x, x_1,x_2, p, q, z)
\langle A \mid \bar{\psi}_q(0)
\frac{\gamma^+}{2}F_{+\sigma}(y_2^-)F_+^{\;\;\sigma}(y_1^-)
\psi_q(y^-)\mid A \rangle\, ,
\label{mod1}
\end{eqnarray}
where $x_3=x+x_1-x_2$ by momentum conservation and $z$ is the
fractional momentum carried by the final quark $\ell_q^-=zq^-$.
The hard partonic part in general has the form,
\begin{eqnarray}
\widetilde{H}^{\rho\sigma}_{\mu\nu} (x, x_1, x_2, p,q,z)
&=& \int \frac{d^4\ell}{(2\pi)^4}
 2\pi \delta_+(\ell^2) \delta(1-z-\frac{\ell^-}{q^-}) \nonumber \\
&&\hspace{1.0in}\times \frac{1}{2}
{\rm Tr}[p\!\!\!/ \gamma_\mu \hat{H}^{\rho\sigma} \gamma_\nu ] \, .
\end{eqnarray}
For the central-cut diagram of quark-gluon rescattering in Fig.~\ref{fig4},
$\hat{H}^{\rho\sigma}$ is given by,
\begin{eqnarray}
\hat{H}^{\rho\sigma}&=&-e_q^2 C g^4 \frac{x p\!\!\!/+q\!\!\!/}
{(xp+q)^2-i\epsilon}
\gamma^\alpha \ell\!\!\!/_q \gamma^\beta
\frac{x_3 p\!\!\!/+q\!\!\!/}{(x_3p+q)^2+i\epsilon}
\frac{1}{(x_1p-\ell)^2-i\epsilon} \nonumber\\
& &\times \frac{2\pi \delta_+(\ell_q^2)}{((x+x_1-x_3)p-\ell)^2+i\epsilon}
g_{\alpha \alpha'} g_{\beta\beta'} g_{\lambda\gamma}
\Lambda^{\beta'\lambda\rho}\Lambda^{\gamma\alpha'\sigma}
\nonumber\\
&=&-e_q^2 C g^4 \frac{(xp\!\!\!/+q\!\!\!/)
\gamma^\alpha \ell\!\!\!/_q \gamma^\beta(x_3p\!\!\!/+q\!\!\!/)}
{(2p\cdot q)^4(x-x_B-i\epsilon)(x_3-x_B+i\epsilon)}
\frac{1}{(1-z)^2} \nonumber\\
& &\times \frac{2\pi \delta_+[2zp\cdot q (x+x_1-x_B-x_L)]}
{(x_1+i\epsilon)(x+x_1-x_3-i\epsilon)}
g_{\alpha \alpha'} g_{\beta\beta'} g_{\lambda\gamma}
\Lambda^{\beta'\lambda\rho}\Lambda^{\gamma\alpha'\sigma} \, ,
\label{mod2}
\end{eqnarray}
where $C=C_A/2N_c$ is the color factor, $\ell$ and $\ell_q=(x_1+x_2)p+q-\ell$
are the momenta of the final state gluon and quark, respectively, and
\begin{equation}
x_L=\frac{\ell_\bot^2}{2p^+q^-z(1-z)}
\end{equation}
is the fractional momentum taken away from the initial state quarks and
gluons by the final state quark-gluon splitting. The three
gluon vertices are defined as
\begin{eqnarray}
&&\Lambda^{\beta'\lambda\rho}=g^{\beta'\lambda}(2\ell-x_1p)^\rho
-g^{\lambda\rho}(\ell+x_1p)^{\beta'}+g^{\rho\beta'}(2x_1p-\ell)^\lambda
\nonumber\\
&&\Lambda^{\alpha'\sigma\gamma}=g^{\alpha'\sigma}(2x_2p-\ell)^\gamma
-g^{\sigma\gamma}(\ell+x_2p)^{\alpha'}
+g^{\gamma\alpha'}(2\ell-x_2p)^\sigma.
\end{eqnarray}
In the above, we have used gluon propagators in the Feynman gauge. Since
the initial gluons only have transverse components in their polarization,
we can also replace the summation of the final state gluon's polarization
tensor as $\sum_i \epsilon^\lambda(i)\epsilon^{*\gamma}(i)=-g^{\lambda\gamma}$.

One can carry out the integrations over $x$ and $x_1$ in Eqs.~(\ref{mod1})
and (\ref{mod2}) by contour integration. There are four possible poles
in the denominator of Eq.~(\ref{mod2}) from quark and gluon propagators.
Different choices for the pair of poles represent subprocesses with
different kinematics. Here we choose the poles at $x=x_B,x_3=x_B$
which correspond to the double hard scattering.
In this case the momentum fraction carried by the initial
gluon is $x_1=x_L$ which scatters with the propagating quark
that has initial momentum $x_B p+q=[0,q^-,0_\perp]$. Therefore, in this
case of double hard scattering, the hard part can be written as
\begin{eqnarray}
\hat{H}^{\rho\sigma}&=&(x_Bp\!\!\!/ + q\!\!\!/)
\frac{1}{4q^-}{\rm Tr}[\gamma^- \hat{H}^{\rho\sigma}] \, .
\end{eqnarray}
Since the contributions from soft gluon radiation  $(z \rightarrow 1)$
is dominant in fragmentation process,
we can take the helicity amplitude approximation,
$\ell\!\!\!/_q \approx z q^- n\!\!\!/$ in the trace of the hard part.
We have then,
\begin{eqnarray}
&& \int dx \frac{dx_1}{2\pi}\frac{dx_2}{2\pi}
e^{ixp^+y^- +ix_1p^+y_1^- -ix_2p^+y_2^-}
\frac{d_{\rho\sigma}}{2x_1x_2} \hat{H}^{\rho\sigma} \nonumber\\
&&\hspace{0.2in}=-e_q^2 \frac{Cg^4}{4}2zq^-
\frac{(x_B p\!\!\!/+q\!\!\!/)2\pi 2p\cdot q}
{(2p\cdot q)^4(1-z)^2x_L^4}
{\rm Tr}\left[\gamma^+\gamma_\alpha l\!\!\!/_q \gamma_\beta\right]
\frac{d_{\sigma\rho}}{2} \Lambda^{\beta\lambda\rho}
\Lambda_\lambda^{\,\,\,\alpha\sigma}\nonumber\\
&&\hspace{0.2in}=e_q^2 Cg^4\frac{z^4(1+z)^2}{\ell_\perp^4}
[(x_B p\!\!\!/+q\!\!\!/)2\pi 2p\cdot q] \, .
\end{eqnarray}

For the soft radiation approximation $z\approx 1$, one obtains in this
light-cone gauge approach,
\begin{eqnarray}
\frac{dW^{(2)}_{\mu\nu}}{dzd\ell_\perp^2}&=&\sum_q H_{\mu\nu}^0(x_Bp,q)
\frac{\alpha_s^2}{\ell_\perp^4}\frac{C_A}{2N_c}\frac{4}{1-z} T^A_{qg}(x_B,x_L) \, ,
\end{eqnarray}
which is the same as the result for double hard scattering in the
covariant gauge \cite{gw-twist}.

In the calculation of semi-inclusive DIS, the transverse momentum $\ell_\perp^2$ 
enters as another scale in additional to $Q^2$ of the virtual photon.
The twist-four contribution to the semi-inclusive spectra at the order of
$\alpha_s$ as a result of the induced gluon bremsstrahlung is suppressed 
by $\alpha_s/\ell_\perp^2$ as compared to the leading twist result,
\begin{eqnarray}
\frac{dW^{(0)}_{\mu\nu}}{dzd\ell_\perp^2}&=&\sum_q H_{\mu\nu}^0(x_Bp,q)
\frac{C_F}{2\pi}\frac{\alpha_s}{\ell_\perp^2}\frac{1+z^2}{1-z} f_a^A(x_B) \, .
\end{eqnarray}
The collinear expansion in this kind of inclusive processes is only valid
when $\ell_\perp^2$ is much larger than the average transverse momentum
of the initial parton or the quark transverse momentum broadening which
is given by the jet transport parameter $\hat q$ [Eq.~(\ref{ptbrod1})].
We have also neglected contributions proportional to $\ell_\perp^2/Q^2$
for $\ell_\perp^2\ll Q^2$. For small values of $\ell_\perp^2\ll \hat q R_A$
the twist-expansion method will fail for the semi-inclusive processes and
one needs to regularize the divergency of the semi-inclusive spectra.
However, the LPM interference between double hard and soft rescattering 
processes suppresses the induced spectra for small $\ell_\perp^2 R_A/2q^- \ll 1$.
Therefore, the final result in the collinear expansion will be a good 
approximation and insensitive to the regularization for large initial
quark energy $q^- \gg \hat q R_A^2$.

%%%%%%%%%%%%%%%%%%%%%%%%%%%%%%%%%%%%%%%%%%%%%%%%%%%%%%%%%%%%%%%%%%%%%%%%%%%%%%%%%%%%%%%%%%%%%%%%%%%%%%%%%%%%%%%%%%%%%%%%

\section{Summary }

In this paper, we have investigated the gauge invariance of
the leading twist-four contribution to the semi-inclusive cross
section of DIS off a large nucleus due to multiple
parton scattering in the framework of generalized collinear
factorization.

We first proved the general equivalence of the hard parts of double
scattering in light-cone and covariant gauge, using a set of identities
for hard partonic processes which were derived from Ward identity and
equation of motion. This equivalence hold for double parton scattering
in the twist expansion to  all orders in $\alpha_{\rm s}$.
We also give two specific examples to demonstrate the equivalence
explicitly. It is easy to see that our proof can be directly extended
to other higher twist contributions.
The equivalence of the hard parts in different gauge at any twist level
can be proved by connecting the contributions of the transverse and
longitudinal components of the gluon using Ward identity as we did.

We have also demonstrated explicitly the gauge invariance of higher
twist contributions in the calculation of semi-inclusive DIS cross
section in the lowest order and the next-leading-order with induced
gluon emission. We pointed out the importance of the gluonic poles in
the calculation of the higher-twist contributions and that interaction
with transverse gluons only leads to higher-twist inclusive cross sections 
which are power suppressed by $1/Q^2$. The leading contribution to transverse
momentum broadening comes from interaction with only the unphysical
gluons.

\begin{center}
{\bf ACKNOWLEDGMENTS}
\end{center}

This work is supported  by the Director, Office of Energy
Research, Office of High Energy and Nuclear Physics, Division of
Nuclear Physics, of the U.S. Department of Energy under Contract No.
DE-AC02-05CH11231 and National Natural Science Foundation of China
under Project No. 10525523.

\end {document}